# Einstein and Besso: From Zürich to Milano.

Christian BRACCO

Syrte, CNRS-Observatoire de Paris, 61, avenue de l'observatoire, 75014 Paris
CRHI, EA 4318, Université de Nice-Sophia Antipolis, UFR LASH, BP 3209, 98 Bd Hériot, 06204 Nice Cedex 3. E-mail : *cbracco@unice.fr*

**Abstract:** The 1896-1901 Milanese period is a key one to understand Einstein's training background. When he was a student at the ETH in Zürich (the Swiss Federal Polytechnic in Zürich) from 1896 to 1900, he would make regular trips back to Milan to stay with his family who was involved in the development of the electricity industry in northern Italy. Between 1899 and 1901, he would meet his faithful friend and collaborator, Michele Besso in Milan on a regular basis. Given their relationship, the 1899-1901 Milanese period therefore foreshadowed the Bern period later in 1904. In order to specify the circumstances under which Einstein and Besso got the chance to meet, we will show that their respective families did have interconnected social networks, especially through the electricity sector and the polytechnic engineering Universities of Zürich and Milan. The branch of the Cantoni family, on Michele's mother's side, rather ignored by now, played a crucial role: with Vittorio Cantoni, a renowned electrical engineer who had not been previously identified as being Michele's uncle, and Giuseppe Jung, professor at the Milan *Politecnico*. We will also show that when staying in Milan, Einstein, who lived in a well-known Milanese palace in the heart of the city, worked in the nearby rich library of the *Istituto Lombardo, Accademia di Scienze e Lettere* in Brera. The linkage between the scientific observations which can be found in Einstein's correspondence and this library collection sheds new light on his scientific work, in particular on his original thesis.

## 1. Introduction.

To date, there have been many biographies of Albert Einstein (1879-1955) (Reiser 1931; Franck 1947; Seelig 1956; Pais 1982; Pyenson 1985; Fölsing 1997, etc.). In addition, several letters from Einstein to his close friends have been released, in particular his letters to his friend Michele Besso, *Albert Einstein, Michele Besso: Correspondence: 1903-1955*, published by Pierre Speziali (Speziali 1979), referred to as EB in what follows, and his letters to Mileva Marić written between 1897 and 1902, published by Jurgen Renn and Robert Schulmann (Renn 1992). Today, the Collected Papers of Albert Einstein documentary edition consists of fourteen volumes and covers the 1879 to 1925 period. [1] The first volume, published by John Stachel (Stachel 1987), referred to as CP1 in what follows, focuses on the 1879 to 1902 period. Among other things, it contains Einstein's letters to his friends Conrad Habicht and Marcel Grossmann, those sent to the professors with whom he applied for an assistant position and Mileva Marić's letters to her friend Helena Savić. His correspondence with Mileva whom he married in 1903, can also be found in this first volume. Even today it still remains the main source for documentation providing an overview of Albert Einstein's scientific concerns between 1898 and 1902.

[1] This edition has been carried out under the *Einstein's papers project* from the *California Institute of Technology* and *Princeton University*, with the archives of the Hebrew University of Jerusalem.

It may be recalled that Albert Einstein entered the ETH [2] in Zürich in October 1896 and graduated in 1900. He then started to complete a first thesis on molecular forces with Professor Alfred Kleiner from the University of Zürich. What is less well known perhaps is that during that same period he spent in Switzerland, Einstein went to Milan during school holidays, for about three months a year, until 1901. During the 1899-1901 period, these stays in Milan took on special importance, since he would meet his friend Michele Besso there, an engineer with whom he would discuss scientific issues almost on a daily basis. The latter was six years older than Einstein. Einstein had met him in Zürich a few years before. Besso, who graduated from the ETH Zürich in 1895, settled in Milan in 1899 where he stayed until October 1901. It was only in January 1904 that they were brought together again in Bern where Einstein was appointed to his first position at the Patent Office as soon as 1902. The 1899-1901 Milanese period which has hitherto been underestimated or ignored altogether, is therefore of particular importance as it foreshadowed the Bern period.

The development of Einstein's ideas is closely related to his friendship with Besso, in particular in Milan between 1899 and 1901, in Bern as from 1904 and in Zürich in 1913. In June 1905, in the acknowledgements of his ground-breaking paper concerning the theory of special relativity, Einstein wrote: *"In conclusion, let me note that my friend and colleague M. Besso steadfastly stood by me in my work on the problem discussed here, and that I am indebted to him for several valuable suggestions."* Later on in 1922, during an impromptu lecture at Kyoto University, while looking back at this period, he said: *"By chance, a friend of mine* [Besso] *living in Bern (Switzerland) helped me* […] *I had various discussions with him. Through them it suddenly dawned on me"* (Einstein 1982). For his part, Besso recalled his presence near Einstein at key moments. For example, in July 1941, at the end of his letter, he alluded to the theory of relativity and commenting on Einstein's quote *"all this can only be conceived by the very person who experienced it"* taken from *The World As I See It*, he wrote: *"In writing this, I do in fact have the feeling I'm falling back into childhood ... 1898, and early 1904, by your side…"* (EB L. 139). [3] In 1947, he noted: *"On my side, I was your public in the years 1904 and 1905; in helping you to edit your communications on the quanta I deprived you of a part of your glory, but, on the other hand, I made a friend for you in Planck* (EB L. 92). Finally, concerning general relativity, at the end of a letter sent to his uncle Marco on June 13th 1913, [4] Besso wrote: *"I will return there* [to Zürich] *to watch my friend Einstein struggle with the great Unknown: the work and torment of a giant, of which I am the witness - a pygmy witness - but a pygmy witness endowed with clairvoyance. "* [5]

In order to gain a better understanding of Albert Einstein's Milanese period and his ties with Michele Besso, the professional and family circumstances under which these two friends got the chance to meet need to be clarified. First we will focus on the bonds the

---

[2] Eidgenössisches Technischen Hochschule (Polytechnikum).

[3] The dotted lines are from Besso.

[4] Marco Besso Foundation, Largo delle Stimmate 1, 00186, Rome, Lazio, Italy. This letter is not reported in the *Collected Papers of Albert Einstein*. This is the only letter left in the correspondence between Michele and Marco. Michele was then working together with Einstein on writing a manuscript consisting of some forty pages and dealing with the calculations of the perihelion precession of Mercury, based on the gravitation theory developed by Einstein in his paper with Grossmann (Sauer 2014).

[5] In this now famous quote, Seelig reported the way the two friends viewed their informal collaboration: « *Later Besso, who for a few terms taught the fundamentals of patents* [in 1916] *at the FIT* [ETH Zürich], *used the following analogy: Einstein the eagle has taken Besso the sparrow under his swing. Then the sparrow fluttered a little higher: "I could not have found a better sounding-board in the whole of Europe", Einstein remarked when the conversation turned one day to Besso"* (Seelig 1956 p 71).

Einstein family fostered with the electrical engineering sector, which played a crucial role in Albert's early career. We will also see that the Einstein family had influential contacts in both scientific and political Italian circles. Concerning the Besso family, we will complete the biography provided by Speziali, with a few specifics concerning the paternal branch and additional information concerning the maternal branch, i.e. the Cantoni branch. This way, we will see that the engineer Vittorio Cantoni, in charge of carrying out the first transmission line of alternate current in Italy between Tivoli and Rome in 1892, was an uncle of Michele's. We will also see that the tie between Albert Einstein and Guiseppe Jung, Michele's maternal uncle through marriage, lasted several years. Finally once again, we will dwell on the letters sent to Mileva, in order to show that when in Milan, Einstein studied in the library of the *Istituto Lombardo Accademia di Scienze e Lettere*, where he was able to find all the science journals he needed for his research, in particular for his doctorate dissertation.

## 2. The Einstein family business, the electrical engineering sector and the Polytechnic schools

The history of the Einstein family is well known, in particular thanks to the numerous biographies about Albert Einstein. Nevertheless, there are still two points we must look more closely at. The first point concerns the 1891 Frankfurt electrical engineering exhibition during which the Einsteins probably established direct contact with Heinrich Friedrich Weber, Albert's professor-to-be at the ETH in Zürich. The second point concerns the ties between the Einstein family and high-ranking individuals from the fields of industry and politics in northern Italy, which led them to move to Milan as from 1894.

### 2.1 The 1891 Frankfurt electrotecnical exhibition, Weber, and Albert heading off for the ETH in Zürich

Albert's uncle, Jakob Einstein (1850-1812) graduated from the Stuttgart Polytechnic School in 1869. As soon as 1879, he became a member of the polytechnic engineers association in Munich, which enabled him to be in touch not only with other engineers concerning technical issues but also with representatives of large companies. The polytechnic school community played an important role for the Einstein family. Jakob was the one who got his elder brother Hermann (1847 - 1902) - Albert's father - involved in the creation of an electrical engineering company in Munich in 1885, the J. Einstein & Co, which employed up to two hundred people. Jakob held seven patents, relating in particular to arc lamps and electrical strain gages. The history of the Einsteins' Munich-based company, viewed within the evolving context of electric companies in Germany, has been told by Nicolaus Hettler in his PhD thesis (Hettler 1996); specific information has also been given by Pyenson (Pyenson 1982).

It is worth noting that the Einsteins took part in the 1891 Frankfurt electrical engineering exhibition. They provided lighting for a number of public places: Milani café, beer shed, labyrinth and shooting range (Steen 1891). They had a stand to present their company which manufactured arc lamps and filament lamps, but also dynamos to generate on-site electricity (fed by steam or gas). Their company was also active in the transmission of electrical current for the supply of DC machines and public lighting. In addition, it produced all kinds of electrical devices. A first conference - a political one - took place in Frankfurt from 27$^{th}$ to 29$^{th}$ August to deal with the situation of municipal lighting and the long-distance electrical power transmission system in Germany. A second conference - a scientific one - followed from 7$^{th}$ to 12$^{th}$ September, with 600 delegates, a third of whom were foreigners. Pyenson mentions that Jakob made a presentation there, along with twenty other participants, which *"described the electrical distribution system that the company had set up in the Munich* *quarter of Schwabing, as well as in the small, northern Italian towns of Varese and Susa"* (Pyenson 1982, p 45).

The Frankfurt Exhibition is famous for the inauguration in August 1891 of the first transmission line (178 km) of alternate current in Europe, from Lauffen am Neckar to Frankfurt. A commission (Prüfungskommission) met there, to compare the electrical systems and devices displayed at the exhibition. It was chaired by the famous German physicist Hermann von Helmholtz, and composed of nine groups. At the origin, it was spurred on by Frankfurt elected officials in 1887, because they wanted to be able to choose between the various electrical lighting systems proposed to the town council. [6] The 1891 commission report appeared in the *Bericht über die Arbeiten der Prüfungs-Kommission* (Offizieller Bericht 1894). The Einsteins' arc lamp system was tested in a comparative study, together with five other systems, by group II dedicated to lighting engineering (p 130). As for Heinrich Friedrich Weber, he was in charge of group IX concerning the transport of energy between Lauffen and Francfort. Within the Prüfungskommission, Weber was therefore aware of the Einsteins' lightning system. He might even have heard Jakob speak during a lecture, and might have known him as well through papers about his patents which had been published in the electrical engineering journals. A bond, perhaps a direct one, might therefore have been created as soon as October 1891 between Weber, Professor of Experimental Physics at the ETH in Zürich and the Einsteins.

This assumption could be confirmed by two elements concerning Albert Einstein's life as a young man. First, after Einstein failed his entrance exam to the ETH in Zürich in October 1895 (at the age of sixteen and a half years), an editorial note in the CP specified that he was nevertheless given permission to attend Prof. Weber's lectures *"Yet* [Einstein] *must have done well on the scientific part* [of the exam], *since physics professor H. F. Weber gave Einstein permission to attend his lectures if Einstein stayed in Zürich"*, which was not allowed under regulation, as specified in note 8 (CP1, p 11). This informal proposal might confirm the existence of a bond between the Einstein family and Weber, or that Galileo Ferraris interceded with Weber, in response to the *"piccola raccomandazione privata"* ("*short private recommendation*") Einstein had asked him on 12$^{th}$ August 1895 to the attention of the latter. [7] Second, Fölsing makes it clear that Jakob played a part in Albert's decision to enter the ETH in Zürich and underlines the fact that his books of geometry do appear in the library of Albert Einstein who annotated them. (Fölsing n. 34, §2). Thus, Albert's demonstration of a mathematical theorem on the developable quality of a cylindrical surface is dated 1891-1895 by the publishers on the basis of *"Einstein's assertion that he started to study plane geometry at the age of twelve* [1891]*… He never took a course in solid geometry, but the subject was among the required topics of the ETH entrance exam, which he took in 1895"* (CP 1 Doc. 3). The prospect of Einstein entering ETH Zürich may have arisen as soon as 1891 at the Frankfurt Exhibition when Albert was only 12 years old.

### 2.2 The Einsteins moving to Milan and Pavia

In 1891, the Einsteins had been turned towards Italy for a few years already (Pyenson 1985; Hettler 1996), since they had among other things carried out the lighting for the town of

[6] Proposals mainly based on the DC power supply lighting were opposed by other proposals backing the AC one (Hettler 1996).

[7] This request from Einstein appeared in a letter which belongs to the Galileo Ferraris archives in Milan, discovered by Andrea Silvestri, professor at the Milan *Politecnico*, member of the *Istituto Lombardo*, and at the time manager of the school's historical records. Published on 27$^{th}$ November 2005 as a Sunday supplement to Sole-24 Oro the Milanese newspaper, it was reprinted in the *Politecnico's* paper to pay tribute to Einstein on 1$^{st}$ December 2005, along with a note (Silvestri 2005). This letter is signed *"Alberto Einstein, presso* [from the] *Ing. Einstein Garrone, Pavia."*

Varese in 1887. Their leaving Munich for Italy was a decision Jakob made, as Maja, Albert's sister, remembered it. Jakob had established contacts with engineers from the Turin or Milan polytechnics and with high-ranking individuals from the fields of politics and industry so as to make it possible for the Munich company to be transferred to Pavia and Milan. According to Pyenson, their leaving Munich had been prompted by contract losses in the town lighting system in 1893. It had also been caused by the crisis affecting electricity companies in Germany and by the wave of restructuring that followed, as deemed by Hettler. At the same time, new opportunities had come up in Italy. Indeed, the industry in the north of the peninsula (Piedmont and Lombardy) started to turn to Germany to purchase industrial equipment from them, as war over the tariff issue had been raging against France. The Italian State, who was quickly developing the country's infrastructure in the road and rail transport sectors on credit, was looking for investors. They turned to Germany and Switzerland for its financing (Milza 2014). The Einsteins' company which settled in Italy as from 1894, can be considered as a case study of this period.

Albert Einstein, who was then a fifteen-year old boy, joined his family in Milan in December 1894. He left the Luitpold Gymnasium on his own initiative, and thus stopped attending school for the rest of the year. He was not left on his own though as he seemed *"occasionally even to have helped in Uncle Jakob's design office"* (Fölsing 1997, p 35). This is probably when young Albert became familiar with scientific journals: the weekly German journal *Electroteknische Zeitschrift* and the French journal *La lumière électrique.* Jakob whose name appeared in these journals must have them in his design office: they contained market news, calls for tenders, exhibitions, patents, recent book reviews, technical data, drawing boards of electrical systems, etc. They can also be found in all the Polytechnic Universities Schools. In Italy, *Elettricità*, a Milanese journal and *Elettricista*, a Florentine one, supplemented the German and French journals. The French journal - four volumes of six hundred pages every year - served as a benchmark at the time and it was unique insofar as it presented electrodynamic theories (Maxwell, Hertz, Helmholtz, etc.) through lectures intended for engineers. [8] This is most likely the first time Albert had the opportunity to come into contact with electricity and electromagnetism, from the theoretical and applied angles. It seems he wrote one first paper in summer 1895, while preparing his entrance exam to ETH Zürich, entitled *"On the investigation of the state of the ether in a magnetic field"* (*CP* 1 Doc. 5). This paper he told his uncle César Koch about, confirms the fact that in doing some reading, he had become quite familiar with this subject.

After leaving Munich, the Einsteins settled in Milan in the autumn of 1894 at the *Palazzo Ricordi*, 2 via Berchet, located at the corner of via Ugo Foscolo (*CP* 1, liii n. 19), just steps away from Milan Cathedral. They would divide their time between Milan and Pavia, where their new factory had just been completed. [9] In March 1894, they became partners with the engineer Lorenzo Garrone in order to set up the *Officine Elettrotecniche Nazionali Einstein, Garrone e C.* registered with the Pavia Chamber of Commerce. This company designed *"dynamos, electric DC and AC engines, arc lamps, voltmeters and ammeters, manual and automatic regulators, switches, metering devices, long distance transmission of electricity, etc."* and employed eighty people. [10] At the beginning of 1895, they moved to

[8] As for example the papers of some forty pages that Camille Raveau devoted to Helmholtz's work, as part of *Recherches récentes sur la théorie électromagnétique de la lumière* (Raveau 1893) (Recent Research on the Electromagnetic Theory of Light) published within three months. In particular, a paper concerning Helmholtz' Least Action Principle in electrodynamics follows a paper presenting Jacob's arc lamps device (p 320).

[9] Information about Einstein's life in Italy is given in [Silvestri 2005] and [Fregonese 2005], as well as a detailed bibliography.

[10] http://www.pv.camcom.gov.it/index.phtml?Id_VMenu=568

Pavia, via Ugo Foscolo, where Hermann rented the house of the renowned poet (*CP* 1, liv n. 20) while obviously keeping his 2 via Berchet address in Milan for the company's office. [11] In 1896, the company had an office in Milan at 41 via Manzoni in the *Borromeo-d'Adda* Palace located six hundred feet from the *Politecnico* (located at *Piazza Cavour* at the time), and offices in Pavia and Turin, as indicated in the catalogue whose cover was reprinted in (Bevilacqua 2005). Following some problems which arose in Pavia concerning the purchase of a concession to use water from the *Ticino River* and contracts for the lighting of private buildings (Bevilacqua 2005), the company was absorbed in September 1896 by the *Carlo Monti e C* [12], which had been registered at the Chamber of Commerce of Milan in May 1895, and whose headquarters was also located at 41 via Manzoni. After it was acquired, the Monti Company became the *Carlo Monti e C. e Rosati*. [13] Even if it looks like he kept on working within the new company for a while, Hermann Einstein was not one of the new company's three managers though. Note that the merger was completed under the leadership of some prominent figures. Sections 12 to 14 of document n. 310 of Milan Commercial Court, dated 2nd November 1896, regarding the new company stipulated that the loss of a quarter of the business value would automatically result in its liquidation under the auspices of Aureliano Scrosati [14], Giuseppe Colombo (who will be discussed again further on in the report) and Giovanni Battista Pirelli. [15]

### 2.3 Via Bigli 21 in Milan.

The Einstein family moved back to Milan in October 1896, at 21 via Bigli. It was not until 23rd February 1899 that Hermann, as the only one person with signing authority, registered a new company with the Milan Chamber of Commerce, whose transfer from Pavia he justified by a '*a greater opportunity*'. [16] The production workshops were located at 160 via Antonio Lecchi and its executive office at 21 via Bigli. [17]

The Einsteins' address in Milan is that of a famous palace located a few steps away from the Scala theatre. The building was sold in 1879 by Gian Giacomo Poldi Pezzoli to Prince Luigi Alberico Trivulzio who was thus the Einsteins' landlord. [18] Countess Clara Maffei, a famous public figure in Milan intellectual and political life, used to rent a flat there, after separating from Andrea Maffei, from 1849 until she died in 1886. Her 'salon', one of the most famous in Milan, attracted the patriotic and reform-minded political elite of the Lombardy region, renowned artists for instance, such as Giuseppe Verdi or entrepreneurs, such as Giuseppe Colombo (Cambria 1982). In a letter her friend Carlo Tenca sent to the

---

[11] As confirmed by the address on the company card that was enclosed to the letter Albert Einstein sent to Ferraris (Silvestri 2005).

[12] Carlo Monti graduated from the Milan *Politecnico* in 1882. The school yearbook [Bolletino 1927] presented him as *direttore dell'Officine illuminazione della Società Edison* and managing director of *Ing. Monti e C.*

[13] *Ing. Carlo Monti e C. e Rosati*. The capital was increased to ITL 250,000 when a partnership was formed with Ferdinando Rosati. An advertisement under this company's name was placed at the end of the first volume of the *Atti della Associazione Elettrotecnica Italiana* (*AEI*) in 1898. Ultimately, it was also liquidated in May 1900.

[14] Aureliano Scrosati was a renowned lawyer who counted among his clients several industrialists (among whom Carlo Erba in particular)

[15] Giovanni Battista Pirelli (1848-1932), a graduate of the Milan *Politecnico* (The Polytechnic University of Milan) in 1870, founded the first Italian tyre industry in Milan in 1872. In 1880, he launched the manufacturing of insulated electrical cables (Gatti 2008).

[16] The creation of *Einstein & Co* was recorded in the register of the Chamber of Commerce in Milan on 4th July 1899 (N. 15739).

[17] In 1889, Hermann built an electricity-producing plant in Canneto sull'Oglio in Mantua and in April 1900 he built another one on the Isola della Scala on the commune of Verona (*CP* 1, lv n. 29); in September 1900, he took Albert to visit them (*CP* 1 Doc. 74) probably to prepare his son to take over.

[18] Cadastral data from the city of Milan.

Countess on 22 October 1882, he described to her with a great amount of enthusiasm the creation of an electricity company and pointed out the fact that the first house to be lighted by electricity would be the house of Giuseppe Colombo, its founder. On 26$^{th}$ December 1883, Colombo and his team (to which Carlo Monti belonged) designed the electric lighting for *La Gioconda*, an opera by Almicare Ponchielli and *Flick e Flock*, a ballet by choreographer Paolo Taglioni (Colombo 2014; Righi 2014) thanks to the electricity produced by the dynamos of the *Santa Ragonda* plant.

The flat the Einsteins occupied at this address was composed of *"eleven rooms with two sumptuous reception room communicating through a large archway"* (Bevilacqua 2005), which may be considered as surprisingly luxurious when we know the family had to face one setback after another in the industrial sector, which made Maja say that at this time *"The family had hardly anything left"* (*CP* 1, p xvii). Therefore, it seems that, as soon as they arrived in Milan, the Einstein were in relation with prominent figures who provided them which such standards of living. In Maja's memories, there is one which proves the existence of these influential relationships: *"The hot summer of 1895 was spent in Airolo on the Gotthard, where young Albert gained a fatherly friend in the Italian minister Luzzatti, who happened to be staying there"* (*CP* 1, lxv and n. 65). Luigi Luzzatti (1841-1927) was a professor of statistics and political economics at the Milan *Politecnico* when it was founded in 1863. [19] He was a colleague of Colombo and he succeeded him for a transitional period on two occasions, first in 1892, as Finance Minister of the Kingdom of Italy, then in 1896, as Treasury Minister.

### 2.4 Two essential actors: Giuseppe Colombo and Galileo Ferraris.

Giuseppe Colombo, State Secretary and Speaker of the Chamber of Deputies, appointed twice Minister, was the rector of the *Politecnico* from 1897 until he died in 1921. He was also a town councillor in Milan.

Colombo was a leading actor in the Italian industrial landscape in the field of electricity. After studying at the Pavia University, he began his career in 1857 in the *Società d'incoraggiamento d'Arti e Mestieri*, where he was teaching descriptive geometry, then he was appointed professor at the Milan *Politecnico*, when it was founded in 1863 and where he was teaching mechanics and industrial machinery (Lori 1941). He had been a member of the *Istituto Lombardo Accademia di Scienze e Lettere* since 1862, its Vice-President two times (1890-1891; 1894-1895) and its President three times (1892-1893; 1896-1897; 1920). Colombo is the founder of the *Comitato promotore per l'Applicazione dell'Energia Elettrica in Italia* (1881). This committee was the prelude to the setting up in January 1884 of the first Italian electricity company, the *Società generale italiana di elettricità sistema Edison*, over which he would preside.

On his part Galileo Ferraris (1847-1897), professor at the *Museo Industriale di Torino* as from 1873, [20] helped develop many technical innovations, among which the first turning-field electric motor in 1885. He also carried out an in-depth study of the Gaulard & Gibbs's transformer and suggested the power plants generating alternate current use it in order to carry the electrical current. He was also a recognized expert in various types of lamps for electric lighting. In 1887, he was one of the prominent figures appointed to lead the Frankfurt commission, responsible for deciding on tenders (mentioned above), and took part in the *Prüfungskommission* who met there in 1891. He served as vice-president of the Congress which was held at this exhibition. At this exhibition, he also led a large delegation of Italian engineers from the *Politecnic* Italian schools. He was a town councillor in Turin and was also elected Senator.

[19] The *Politecnico*, founded in 1863, whose name at the time was *Regio Istituto Tecnico Superiore*, graduate about sixty engineers every year, among whom ten electrical engineers or so as from 1887. It was founded with the help of the *Società d'incoraggiamento d'Arti e Mestieri*, the *Istituto Lombardo* and the *Osservatorio di Brera*.

[20] In 1906, the Museo Industriale di Torino, a school modeled on the *École des Arts et Métiers* in France, merged with the *Reale Scuola per Applicazione degli Ingegneri di Torino*. This new school was renamed the *Politecnico di Torino*.

Following their visit to the Paris Exposition of 1881, where Thomas Edison had given his demonstration, Colombo and Ferraris became the promoters of electric lighting in Italy. Colombo had been negotiating with Edison in the US to get an exclusive license for Italy on the 'Edison's system' of incandescent light bulbs, the electric power supply and distribution systems. At his instigation, Milan became the first city to be lighted by electricity in Europe. In 1897 when Ferraris died, Colombo took over management of the *Associazione Elettrotecnica Italiana* (AEI), an association they created officially on 26th December 1896. [21] This assembly brought together entrepreneurs, academics, professional traders and about five hundred members, divided into six sections (Turin, Milan, Genoa, Rome, Naples, Palermo) one-third of them in Milan. The Einsteins' partner in Pavia was Lorenzo Garrone, a collaborator of these two prominent figures. He was a civil engineer, graduated from the *Reale Scuola per Applicazione degli Ingegneri di Torino* in 1881 and director of the *AEI* Turin section in 1900 (Gatti 2007-2009).

We can wonder whether the Einsteins could have managed to settle in Pavia and Milan without entering into direct contact with Colombo and/or Ferraris. Let us note that while it is likely the Einsteins were in relation with Colombo in Milan, so had they been before with Ferraris: Jakob Einstein wrote to him twice on 26th January and 8th March 1888, about a dispute with Italian engineers concerning the public lighting in Varese. [22]

## 3. The Besso-Cantoni family, the financial community and the Polytechnic Schools.

Michele Besso's uncles held prominent positions in the financial, industrial and scientific sectors, and they were involved in the development of electrification in Italy. The community of the *Politecnico* engineers also played an important part in Michele's family. This information is essential to understand Michele Besso's choices and the circumstances under which he met with Albert Einstein. The biography of the Besso family and Michele's life are indeed partly documented in Speziali's introduction, yet Michele had two uncles who played a remarkable role which has been ignored or undervalued: Vittorio Cantoni and Giuseppe Jung. But first of all, we will focus on Michele's family on his father's side, which will provide some more information about his life.

### 3.1 The Besso family.

Michele Angelo Besso was born in Riesbach near Zürich on 25 March 1873 and according to Speziali, his parents, Giuseppe Besso (1839-1901) and Erminia Cantoni, settled in Trieste - his father's hometown - in 1879. His father had an executive position in the *Assicurazioni Generali*, after managing a subsidiary of this same company in Zürich. Giuseppe's brothers were Marco, Beniamino and Davide, thus Michele's uncles. Davide Besso (1845-1906) was appointed professor of mathematics in 1871, when *Leonardo da Vinci*, Rome Technical Institute located in the Palazzo Cesarini and Borgia, was founded. Then, he was a professor in Modena. He was the founder of the *Periodico di Matematiche*, a journal designed for teaching mathematics in secondary education. Beniamino and Marco had significant responsibilities that need to be mentioned.

[21] The *Atti della Associazione Elettrotecnica Italiana* were edited as from 1898 (approximately 250 pages every year with papers, lecture announcements, technical drawings, chapter updates, General Assembly, etc.).

[22] These letters written on the behalf of the *Elektrotechnische Fabrik J. Einstein and Comp*., are listed in (Gobbo 2005) together with a biography of Ferraris.

### 3.1.1 *Beniamino Besso.*

Beniamino Besso (1844-1907) was an engineer, who entered in 1869 as a non-resident member the *Sociétà degli Ingegneri e degli Industriali di Torino* (founded in 1866). The following year, he was an engineer at the *Officio Studi e progetti delle Ferrovie dell'Alta Italia*. In 1871, he wrote a small illustrated handbook entitled *Il Cenisio, illustrato e descritto* (Besso 1871 a). This small handbook which presented the technical construction of the Fréjus tunnel (Mont Cenis), was published on the very year the tunnel was inaugurated, on 17th September 1871. Beniamino Besso also wrote a book about *Electricity and its Applications* (Besso 1871 b). Speziali presents Beniamino as the director of the Sardinian railways. As the development of the railways played a strategic role in the newly unified Italy (the kingdom of Piedmont-Sardinia being largely instrumental in leading the Italian unification movement), Beniamino had a key political role. [23]

He was also the author of a famous popular science book, re-edited six times within ten years, entitled *The Great Inventions and Discoveries* (Besso 1864). It was in this book that Michele probably learned to read. Indeed, Speziali reports that Michele was *"a precocious child and curious of everything"*, who *"learnt to read when he was about five, in a book which revealed* [to him] *almost everything you could know 90 years ago about physics, technology and astronomy without resorting to mathematical formulae"*. Actually, the first edition of *the Great Inventions* dated back exactly 89 years before Michele Besso's quote in 1953.

Beniamino's wife, Amalia Goldmann Besso, was a renowned painter, to whom the online *Treccani* encyclopedia pays tribute in a paper. It states that she moved to Rome with Beniamino in 1883. An exhibition entitled *Artiste del Novecento, tra visione e identità ebraica,* was dedicated to her by the city of Rome in 2014. [24]

According to Speziali, when Michele enrolled in the University *La Sapienza* in Rome in March 1891, he went to stay with Beniamino and Amalia. His registration file in the University Archives in Rome reads that he enrolled on 24 March 1891 in a first-year mathematics programme as 'future engineer' producing proof that he had completed his *'maturità'* [A-levels] *"awarded on 9th July 1890 by the Ginnasio comunale superiore di Trieste, and recognized as equivalent by the faculty."* [25] He earned outstanding grades at University: 28/30 in projective and analytic geometry with del Re; 30/30 in algebra avec Biolchini; 28/30 in experimental physics with Blaserna who was the Dean of the Faculty of Science at the time. [26] However, Michele did not continue with his studies in Rome, where he could have specialized at the *Scuola d'applicazione per gli Ingegneri in Roma,* founded in 1873 and run by Luigi Cremona, who had previously been a professor at the Milan *Politecnico*. We will see further on the possible reasons which led him to join the ETH in Zürich in 1891, from which he graduated in 1895. Once again, he earned the highest grades (6/6) in experimental physics and mathematics in the classes given by Weber and Hurwitz.

---

[23] It is worth noting that the city of Modane, marking the entrance of the tunnel on the French side, was equipped with public lighting in 1885. The entrance on the Italian side, in Bardonecchia, is less than 30 km away from Susa, whose electrification was implemented by the Einsteins around 1887. Lorenzo Garrone contributed to implementing the lighting of Bardonecchia (Sanesi 1982) before partnering with the Einsteins in Italy. Let us also remember that it was in Airolo, at the Italian end of the Gotthard railway tunnel that the Einsteins got the chance to meet the Italian Finance Minister Luzzatti in 1895. So obviously, the Einsteins had connections with the railway community. It is also worth noting that Galileo Ferraris and Beniamino Besso must have known each other. When he was a student at the *Museo Industriale* in Turin, Ferraris visited the tunnel construction site in 1868 and in 1870 attended the banquet in honour of the engineers who had built the Fréjus tunnel (Gobbo 2005), organized by the *Società degli Ingegneri e degli Industriali* (of which he would be a member eventually).

[24] The paintings lent by the Marco Besso Foundation, depict Amalia's journey around the world in 1912 with her nephew Salvatore, Marco's son.

[25] Speziali reports that Michele had been suspended from school when he was 16 for, together with a classmate, circulating a petition against his maths teacher whom he considered incompetent.

### 3.1.2 *Marco Besso.*

In 1909 Marco Besso (1843-1920) was president of the *Assicurazioni Generali*, the largest insurance company in Italy. He thus played a central role in the Italian financial world, as his correspondence testifies. The latter, whose record is kept in his foundation, [27] is composed of many letters, both official and friendly, from Finance Minister Luigi Luzzatti (with whom the Einsteins were in a relationship as well, as previously noted). There are also a few letters sent by Salvatore Majorana (the father of the physicist Quirino and the grandfather of the physicist Ettore) who was then Minister of Agriculture, Trade and Industry. These letters, dated May 1879, were about the setting up of mutual assistance societies for workers. Indeed companies, such as the one the Einsteins had in Pavia, were expected to set up a fund to provide this type of aid. Therefore, the tie between the Einsteins as entrepreneurs and the Bessos through the *Assicurazzioni Generali*, could be a useful avenue to explore further. Marco Besso was to be the president of several other companies and serve in several Boards of Directors, in the financial sector, but in the industrial electricity sector too. *“As a figurehead”* of the *Generali*, Marco also went down in the history of the electrification of Italy for having raised the capital needed to found, in 1891, the *Società veneta di telefoni et di elettricità di Venezia*, which, together with the development of the telephone, challenged the monopoly of the local company in electric lighting (Mori 1992, p 269). He married Ernesta Maurogonato (1853-1916) who was the daughter of Bersabea Ascoli, the latter being the sister of Graziadio Ascoli, the renowned linguist (1829-1907), who would call Marco *“my dearest nephew”* in his letters. [28] On a proposal by the Ministers, Marco was appointed a Knight then an Officer and a Grand Officer of the Crown of Italy.

### 3.2 *The Cantoni family.*

Michele’s mother, Erminia Cantoni, was Angelo Cantoni (born in Vicenza on 10$^{th}$ September 1825 – 21$^{st}$ October 1887) and Elisa Maroni’s daughter. Speziali gives no information on this branch of the family except that they came from Mantua. Yet, Erminia had two brothers, Vittorio and Tullo, and three sisters Bice, Maria and Emma. While Vittorio Cantoni was known through his active work in the electrification of Italy, he had not been related to Michele’s family yet. It is worth giving bibliographical information about him, for he probably played a major role in the eyes of Michele.

Bice married Giuseppe Jung (1845-1926) in 1879, the professor of mathematics from the Milan *Politecnico*. Jung was presented as Michele Besso’s maternal uncle (*CP* 1 Doc. 94, p 280, n. 8) when Einstein mentioned him in his letters to Mileva as from March 1901. We will see that his relationship with Einstein went on afterwards.

---

[26] Pietro Blaserna was also the founder and director of the Physics Institute located at via Panisperna in 1880. He played a key role in scientific development in Italy after 1870 (Battimelli 2013).

[27] The foundation houses a rich library dedicated to the various editions of Dante’s works and to studies on this author, and to various original documents as well on the region of Lazio (province of Rome), gathered by Marco Besso.

[28] During the unification of Italy, he had defended an Italian language which would take into consideration the linguistic specificities of the various kingdoms.

Tullo Cantoni, born on 27th September 1866 in Vicenza, was Erminia, Bice and Vittorio's youngest brother. He completed his final third year of law studies at the University of Bologna in 1888, after studying at the University of Rome. [29] Tullo, who also inherited the title of Count of Mamiani della Rovere, was mayor of the town of Arona from 1914 to 1917. [30]

Maria (1855-1940) married her cousin Achille Cantoni (1848-1914). In 1905 Arrigo, (1877-1953), one of their five children, took a degree in architecture at the *Reale Scuola per Applicazione degli Ingegneri di Torino*. It was propably he who in 1912 submitted a project for the new train station in Milan. After the First World War, he became teacher of experimental physics in secondary school and in particular, he wrote *L'esperimento fisico nelle scuole medie*, published in 1938 by Hoepli, an editor from Milan. It should be noted that Einstein met him in 1922 as reported by Besso: *"The latest news I had of you was given to me by my cousin Arrigo Cantoni who was greatly pleased by your friendly welcome* " (*EB* Doc. 65).

Lastly, Emma (27th September 1966 – 1946), married Eugenio Norsa (1856-1933).

We note that Michele's brothers and sisters were named after their uncles and aunts: Vittorio Beniamino, Marco Tullo, Luisa Margherita (which are the first names of two daughters of Giuseppe Jung). In addition, Marco Besso archives in Rome own the photography reproduced in the Annex. We see here that the paternal and maternal branches of Michele's family are reunited. They were brought together in Arona in the villa Cantoni, one of the most beautiful on the shores of Lake Maggiore (Lodari 2002, p.115-117). Vittorio Cantoni had this villa built in the late 1880s. [31]

### 3.2.1 *Vittorio Cantoni.*

Vittorio (30th August 1857 in Milan – 1930) entered in 1874, at the age of 17, the ETH in Zürich, which he graduated from on 23rd March 1878. He married Albertina Solal in 1900. His student file at the ETH contained an important piece of information: Michele's father, Giuseppe Besso, was his contact person in Switzerland. Vittorio was probably staying at Michele's parents while studying and when he left Zürich, Michele was 5 years old. Vittorio entered the ETH Zürich mechanical engineering section but the following year he changed for the civil engineering section. It is worth noting that he passed Weber's physics exams with the highest score possible of 6 out of 6 during his second year in 1875-1876. [32] Back in Italy, Vittorio graduated from the Milan *Politecnico* on 9th September 1879. He was able to attend the classes of Giuseppe Colombo, who had just finished writing his famous electrical engineering handbook for engineers (Colombo 1877-1878) published by Hoepli, the renowned editor from Milan. [33]

Very soon, Vittorio assumed significant responsibilities. In 1886, he installed a Gaulard-Gibbs transformer for the construction of the hydroelectric power plant in Tivoli, near Rome. He was then appointed chief engineer of the first electricity line between Tivoli and Rome – first AC transmission line in Italy (28 km) – designed to transport the electricity produced by the waterfalls on the way from Tivoli to Rome and provide city lighting

[29] Archivio storico, università degli studi di Bologna, Fascicolo degli studenti n. 923, in 1888.
[30] Tullo lived in the villa Cantoni with his wife Irma Finzi and their children. His son, Vittorio Angelo, born in 1899 and Irma both disappeared on 15 September 1943 « *vittime del razzismo nazista* » (victims of Nazi racism) as specified on the war memorial in Arona.
[31] Angelo's father had bought this piece of land from the Borromeos in 1873. The architecture and the garden of this villa have been described by Paolo Cornaglia in (Lodari 2002, p 115-118).
[32] Vittorio Cantoni, Anmeldung zur Aufnahme in das Eidgenössische Polytechnikum.
[33] This book is considered the standard reference and an extended version of four volumes is still being published today.

(Marcillac 1893). This line was inaugurated on 4th July 1892 before 500 leading figures. Vittorio worked toward its implementation with Guglielmo Mengarini, professor at the Engineering School of Rome. [34] Mengarini was, just like Galileo Ferraris, a member of the *Prüfungskommission* in Frankfurt and in 1896 he would be the Vice-President of the AEI together with Colombo, before becoming its President in 1903. Therefore it is almost certain that Vittorio was part of the delegation of engineers who accompanied Ferraris to Francfort and that he had the opportunity to speak directly to his former professor Weber about technical matters related to the installation of an alternate current power line (since Weber oversaw the work of the *Prüfungskommission* in this sector, as mentioned above).

At that time, Vittorio was an engineer of the hydraulic power company run by Carlo Esterle (1853-1918). The latter who succeeded Colombo in 1897 to lead the Edison company, was appointed member of the Board of Directors of the *Società per lo Sviluppo delle imprese elettriche in Italia* in 1899. [35] This company hired Michele Besso from 1900 to 1901, according to Speziali: *"Indeed, in 1899, Besso went to Milan to work in a company that generates electrical energy over long distances. From 1900 to 1901, Michele was a technical adviser for the Società per lo sviluppo delle industrie* [imprese] *elettriche in Italia, whose headquarters is located in the capital of Lombardy.* [36] *Over these two years, he actively participated in developing safety measures – as imposed by the Italian insurance companies – regarding the electrical installations."* In the light of this, it appears that Michele's entering the ETH Zürich in 1891 and the employment opportunities he found in Milan between 1899 and 1901, had a direct connection with Vittorio.

### 3.2.2 *Giuseppe Jung.*

Giuseppe Jung is Michele's maternal uncle by marriage. The Jung family have been the subject of recent academic studies (De Ianni 2009; Raspagliesi 2012) which provide, among others, the family genealogy. On 4th April 1901, Einstein mentioned Jung to Mileva: *"The day before yesterday he* [Besso] *went on my behalf to see his uncle Prof. Jung, one of the most influential professors of Italy & also gave him our paper. I met the man once before & must admit that he impressed me as quite an insignificant person. He promised that he will write to the most important professors of Italy (physicists), Righi & Batelli, on my behalf, i.e., ask them whether they need an assistant. This is already quite a lot, because he seems to be on very friendly terms with them"* (*CP* 1 Doc. 96). The paper he gave Jung was probably the paper on capillarity, which was Einstein's first paper published in *Annalen der Physik* in March 1901 (Einstein 1901).

It should be noted that Einstein probably got the chance to meet Jung as soon as 1897 in Zürich, during the first international mathematics symposium. [37] It was organized at the ETH Zürich by Einstein and Besso's professors: Carl Friedrich Geiser, Adolf Hurwitz, Ferdinand Rudio, Heinrich Weber, Albin Herzog. It was placed under the auspices of Henri Poincaré for France and Felix Klein for Germany (Décaillot 2008). The names of Michele Besso and

---

[34] Mengarini introduced the first electrical engineering courses at the Engineering School of Rome in 1891, optional at the beginning, in the year when Michele was a student there. Before that, like Ferraris and Colombo, he had also been to the Exposition of Paris in 1881.

[35] Marco Besso was to be the President of this company too in 1914 (Scolari 1967). This company played an important role in mediating conflicts between Italian and foreign industrial groups, as to the sharing of their spheres of influence. When it was set up, some organisations from Milan provided 20% of the funding and an Austro-German group provided the rest. (Pavese 1993).

[36] That does not seem to have been the case: this company was not recorded by the Chamber of Commerce in Milan.

[37] < http://www.mathunion.org/ICM/ >

Giuseppe Jung were on the list of participants and Jung's address was mentioned on it: 9 via Borgonuovo in Milan, about a hundred metres away from the Einstein family.

From the 1873 academic year onwards, [38] Giuseppe Jung assumed the professorship dedicated to graphic statics, left vacant at the Milan *Politecnico* by Luigi Cremona who had been called to run the Engineering School of Rome as mentioned above. [39] He became a correspondent of the *Istituto Lombardo, Accademia di Scienze e Lettere* in 1879, then a permanent member in 1893 and a resident in 1908. So he was also a colleague of Colombo at the *Politecnico* and at the *Istituto Lombardo*.

The list of Jung's publications (Maggi 1927) was quoted in the tribute paid to him in 1927 in the *Rendiconti of the Istituto,* after his death in January 1926. In November 1926, his youngest daughter, Maria Vittoria, donated Giuseppe's personal library, a part of it to the mathematics library of the *Politecnico,* another part of it to the mathematics department library of the University of Milan who recorded the writings in 1935. This large donation which was composed of approximately 450 books and some 2,500 papers (essentially reprints of papers) can be found in details in these libraries' entry registers. Such a large amount of papers appears to indicate that Jung was in contact with many colleagues, thereby confirming Einstein's impression that he was indeed *"one of the most influential professors of Italy"*.

The list of the 2,500 reprints in the donation was, as expected, only composed of papers about mathematics, with one notable exception: the full list of Einsten's publication reprints from 1901 to 1906, with a total of twelve papers. [40] There had been no evidence until then of an ongoing interaction between Einstein and Jung beyond the '*essay*' given to him by Besso as mentioned above. Could it be that Einstein might have kept up the hope of getting a post as assistant in Italy, before obtaining his first post at the University of Bern in 1908 and in this view might have maintained ties with Jung? Unfortunately, based on the information gathered indirectly from a former head of the library at that time, a decision was made in the late 1970s to send these papers to the shredder, considering their poor state of conservation.

Jung was one of the five (and for instance the first on the list) along with Luigi Berzolari, Antonio Federico Jorini, Francesco Gerbaldi and Giulio Vivanti, to propose on 4th May 1922 the appointment of Albert Einstein as a foreign member of the *Istituto Lombardo,* along with Costa Mittag-Leffler, Paul Painlevé and three Italian members (to achieve parity in the appointments), who were mathematicians as well: Guido Fubini, Giuseppe Peano et Gaetano Scorza. The *Istituto Lombardo* has records of the arguments presented for this proposal in its archives: *"<u>Albert Einstein</u> needs no introduction. Even though the general theory of relativity has had precedents just like all theories, there can be no doubt nonetheless that Einstein has had the merit of boldly laying down the fundamental postulates that the laws of physics are invariant in a uniform translatory motion of a frame of reference and the principle of invariant light speed in vacuum. Later on, while the theory whose foundations he had laid was undergoing strong development, the need to expand its scope of application led him to modify his assumptions by stating that the laws of physics are invariant compared to changes in more general frames of reference. The debate this new theory has triggered in the scientific world, the lively curiosity with which the results of experiments are expected, the recognition*

---

[38] Before holding this professorship, Jung used to be a teacher at the *Liceo Parini* (High School Parini) in Milan (previously *Regio Ginnasio di Brera*). He also gave a lecture to the students who wanted a direct entry to the *Politecnico,* without attending the first two years at the Faculty of Mathematics at the University of Pavia.

[39] Jung collaborated in the third edition of Cremona's book on graphic statics (Cremona 1879). This discipline allows solutions to be reached using geometry construction to solve differential equations without resorting to the accurate analytical method and it is particularly adapted to engineering calculations.

[40] Ref. 3400 and from 2766 to 2776 in the *Politecnico* register of entries in 1926. Einstein's papers completed the donation of Jung's reprints, except for the first one concerning capillarity, which was separated from the others.

*achieved even outside the scientific circle, all this shows that this wealth of exceptional achievements is universally attributed to Einstein's views, so that even those who do not agree with him on some particular issue recognize him as a reformer of natural philosophy."* [41] The election took place in July 1922, shortly before Einstein was awarded the Nobel Prize. [42]

## 4. The Einstein-Besso tie before 1904.

We may wonder under what circumstances Einstein and Besso got the chance to meet. We will in what follows give some insights on how to answer this question. We will then focus on the scientific subjects they discussed together, in order to gain a better understanding of the nature of their collaboration which had remained informal but started to grow in Milan in 1899-1901. We will also see that Einstein encouraged Michele to complete a thesis. We will establish that according to Mileva's correspondence, Einstein was working in Milan, in a world-class science library, the *Istituto Lombardo, Accademia di Scienze e Lettere*. Finally, we will examine a possible direct consequence of his bibliographic work on his 1901 doctoral dissertation.

### 4.1 The Einstein-Besso meeting

It is a letter sent by Albert Einstein to Michele Besso on 6th March 1952 which provides initial information on their meeting. Einstein had just been contacted by Carl Seelig who was writing out a biography about his younger years in Switzerland; he sent a few remarks to Michele: *"I know already that the brave Seelig is looking into my childhood days. It must be said that we would discuss scientific matters each day when returning home from the office.* […] *In Bern, I would attend evenings of philosophical readings and discussion with C. Habicht and Solovine on a regular basis, during which we would deal mostly with Hume (in a very accurate German edition). This reading has had some influence on my development, as well as Poincaré and Mach. You're the one who recommended this book to me during my studies, after we met at the home of Mrs Caprotti"* (*EB* L. 182). In his foreword, Speziali believes this meeting occurred *"in 1897, in Zürich"* while *"Einstein was 18 and Besso 24"*. But in his introduction, in a contradictory way, he says *"We were in 1896, in early autumn. Besso, who often went to Zürich, was friend with a family of music lovers, the Hünis, where he played the violin. It was while he was spending an evening there - the exact date could not be ascertained – that* he met Albert Einstein for the first time (italics in the original). *He was then 23 years old, while Einstein who was born on 14 March 1879, was only 17"* (*EB* p xxii).

In a letter he sent to Eduard Lüscher on 23rd September 1943, Michele Besso wrote that his first meeting with Albert Einstein dated back to the year after he had finished school in Aarau. [43] In a letter he sent to Einstein on 10th October 1945 (*EB* Doc. 145), Besso stressed the fact that they had been friends for fifty years, since about 1895, and that this friendship was born from a discussion they had on the wave and particle models of light: *"Isn't that again the starting point* [of the discussion] *we had fifty years ago, between Newton and Huygens ?"*. After the death of his friend, Einstein wrote to Michele's son and sister that

[41] Proposta per la nomina di tre Soci corrispondenti stranieri nella Sezione si scienze matematiche, *Istituto Lombardo*, 1922.

[42] Einstein and Mittag-Leffler were elected with fourteen votes out of sixteen, the other members were elected unanimously.

[43] This letter (Doc. 71-969-3, Einstein Archives, Jerusalem), sent to an 18-year-old first year student at the ETH, gives a simplified version of events. For example, at the beginning of his letter, Michele explains that Einstein, being too young to sit the entrance exam to ETH, had to complete one year in Aarau, which does not exactly comply with the situation we have seen earlier.

*"Our friendship was born when I was a student in Zürich where we met regularly at musical evenings"*.

Albert Einstein mentioned a meeting in the home of Mrs Caprotti. Selina Caprotti, whose maiden name was Hüber, [44] got married in Zürich in 1871 to Carlo Caprotti (1845-1926), director of the large textile mills *Cotonificio Caprotti* in Ponte d'Albiate near Milan, as recorded in the archives online of the Caprotti family. [45] The presence of Michele in the home of Selina Caprotti brings into play Michele's maternal branch. In the book he dedicated to the Cantoni villa mentioned above, Cornaglia pointed out that Angelo Cantoni, Erminia's father, a banker in Milan, was involved in the textile industry, more particularly in the processing of cotton in Legnano (Lodari 2002). If this information is correct, there could also be a family or at least a professional tie between Angelo Cantoni and Eugenio Cantoni (1824-1888) from Gallarate, made baron by Victor Emmanuel III in 1871, and who ran the largest Italian textile company, the *Cotonificio Cantoni*, the first Italian company to be quoted on the Stock Exchange, whose historical headquarters is in Legnano. [46] Thus, the Caprottis and the Cantonis had a relationship. Let us note that Giuseppe Frua (1855-1937) worked alternately in both companies, until he started together with Ernesto De Angeli (1849-1907), the De Angeli-Frua textile printing company. We also note that Carlo Caprotti and Selina Hüber belonged to the same generation as Michele's parents, Giuseppe Besso and Erminia Cantoni, and that both couples got married one year apart in Zürich.

The date when Albert and Michele first met is an element it would be desirable to clarify. Let us recall that Besso's and Einstein's uncles (Vittorio Cantoni and Jakob Einstein) were both polytechnic engineers (from ETH Zürich and Milan for Vittorio, from Stuttgart for Jakob) who had been either Weber's student (Vittorio), or probably in contact with him (Jakob) and that these people might have been able to meet as from 1891 at the Frankfurt Exhibition. Jakob and Vittorio are also connected to Ferraris, himself close to Weber. Moreover it was probably during this exhibition that Jakob introduced his nephew Albert to Ferraris. [47] Jakob, who was a member of the Polytechnic Engineers' Association in Munich, must have maintained relations with the *Politecnico* and the industrial circles in Milan as well. Giuseppe Jung, Vittorio's brother-in-law, was himself a professor at the Milan *Politecnico*. The Einstein and Cantoni families were tied together by the politechnic community and the electricity sector. Since Michele Besso graduated from ETH Zürich in March 1895 while Albert Einstein was about to sit the entrance exam in October, Michele would have been the right person to support Albert while he was preparing his entrance exam to ETH Zürich, and he might have done so as early as 1895. [48]

---

[44] The surname Hüber is mentioned three times in the Einstein-Besso correspondence: for instance, Eugen Hüber *"who designed the Swiss Civil Code"* (*EB* Doc. 201) and Rudolph Hüber, Mathematics teacher in a Gymnasium in Bern (*EB* Doc. 100, n. 2).

[45] Archives online < www.giuseppecaprotti.it>; in particular < www.giuseppecaprotti.it/wp-content/uploads/albero-genealogico-caprotti1.pdf > to see a family tree online of the Caprottis.

[46] As from 1871 Eugenio Cantoni funded a Chair of Economics at the Milan *Politecnico,* which had been held, under a different name, by Luigi Luzzatti in 1863 (*cf* §2.3).

[47] From the very first lines of his letter to Ferraris dated August 1895 (*cf. n.* 7), Einstein thanks the "*Molto Illustrissimo Sig. Professore*" Ferraris for remembering him, as it was reported to him by the "*Ing. Vitali*" following a conversation with Ferraris.

[48] Note that in the autumn of 1901, Einstein himself was the private teacher of a young Englishman named Louis Cahan, at Dr Nuesch's private school in Schaffhouse, in order to help him prepare his entrance examination to ETH Zürich.

**4.2 Some topics covered by Einstein and Besso before 1904.**

We can see in the letters he sent to Mileva from Milan that Einstein was interested in a wide variety of topics: electrodynamics by Helmholtz and Hertz (1898-1899), the "relative motion of matter and ether" in relation with experiments (1898-1901), the Drude model of electrical conduction in relation with thermoelectrics (1900-1901), molecular forces, topic of a first thesis which had not been completed with Alfred Kleiner (1900-1901), the interaction of radiation with matter in relation with the works of Planck and Lenard (in spring 1901), etc.. In addition to scientific papers, Einstein also read several handbooks propounding the contemporary theories (Hertz 1892; Ostwald 1893; Drude 1894; Helmholtz 1895; Meyer 1895; Boltzmann 1896-1898, etc.), most of them focusing on electrodynamics and molecular hypotheses. To him, these readings continuously conjured up criticisms or new ideas, and his letters to Mileva ended with considerations sometimes general, sometimes specific, on these topics.

In Milan, between 1899 and 1901, Albert would discuss these topics with Michele on a regular basis, as he wrote to Mileva in the late summer of 1900: *"I am spending many evenings at Michele's"* (CP1 Doc. 74). These meetings were an opportunity for extensive discussions on physical issues. Indeed, as of early April 1901, Einstein wrote to Mileva: *"Yesterday evening I talked shop with him with great interest for almost four hours. We talked about the fundamental separation of luminiferous ether and matter, the definition of absolute rest, molecular forces, surface phenomena, dissociation. He is very interested in our investigations, even though he often misses the overall picture because of petty considerations. This is inherent in the petty disposition of his being, which constantly torments him with all kinds of nervous notions*" (CP1 Doc. 96). These discussions were important to Einstein who in October 1900 told Mileva of Michele's *"truly outstanding intelligence"* (CP1 Doc. 79).

The copious correspondance between Einstein and Besso amounts to 215 letters which have been found by Speziali, who believes a hundred of them are still missing. The first letter was sent by Einstein in January 1903, and Michele replied to it as of early February from Trieste (the second listed letter sent by Besso is dated 1911). This letter of 1903 is thus by now the unique testimony of the specific and technical nature of the first years of collaboration. In three pages, Michele described, following a request by Einstein concerning molecular dissociation (one of their topics in Milan) a list of six scientific papers explicitly referred to, with comments in writing, graphs and value tables. It is likely that the two friends collaborated in a similar way on other scientific topics and that a significant aspect of their work involved not only discussing freely scientific topics, but also compiling an extensive bibliography, which is the core of all research work.

This bibliographic search was not limited to recent science journals, but it covered scientific publications too. Besso lent books to Einstein, since in April 1901, he wrote to Mileva he had "*studied electrochemistry and chemical reactions from Michele's 'Ostwald'*" (CP1 Doc. 97) and at the end of 1901 from Schaffhouse, he wrote: *"Michele gave me a book on the theory of ether, written in 1885.*" Giuseppe Jung, Michele's uncle, might have played a role too in transmitting these books. Thus, Einstein read Eduard Heine's book (Heine 1878-1881) in September 1900, concerning heat transfer in a cylinder (CP1 Doc 76 et CP1 p 262). He wrote to Mileva *"By now I have studied the entire Boltzmann and a part of spherical harmonics, in which I have now even got quite interested. Beggars can't be choosers"* then sent her back to Heine in May 1901, for her to complete her ETH degree. This book can be found in Jung's personal library. [49]

### 4.3 Michele Besso's thesis.

For his part, Einstein also worked for Michele, as he explained in the post-scriptum of a letter he sent to Mileva, dated 30th August or 6 th September 1900: *"I am investigating the following interesting problem for Michele: How does the radiation of electric energy into space take place in the case of a sinusoïdal alternate current ? About the amplitude of the waves produced as a function of the frequency of vibration, etc."* (CP1 Doc. 74). On 3rd October 1900, one month later, he wrote to Mileva about Michele *"I tried hard to prod him into becoming 'Dozent', but I doubt very much that he'll do it. He simply doesn't want to let himself and his family be supported by his father, this is after all quite natural"* (CP1 Doc. 79). Einstein, who then began to write his own thesis about molecular forces with Kleiner, obviously prodded Besso into starting his thesis work. Michele did mention he had started a thesis in a letter sent to Einstein in 1928: *"The last two times we met, you reminded me of my doctoral thesis which in fact was never completed, – for long periods I have been thinking of the many more ties, of all different kinds, that bind us together"* (EB L. 92). If the work carried out by Einstein on behalf of Michele mentioned above was related to Michele's thesis, then the latter might have worked on wireless transmission, i.e. the propagation of radio waves, a booming topic at the time, a field in which the Italian engineers and scientists Guglielmo Marconi (1874-1937) and Augusto Righi (1850-1920) had been pioneers. [50]

In the years 1895-1899, Besso had probably little opportunity to devote time to research, as he was appointed to his first position in 1896 at Rieter in Winterthur, an industrial machinery manufacture, met his wife Anna Winteler through Einstein in June 1897, had his son Vero in 1898, changed firms and moved to a different country to settle in Milan as from 1899. In addition to that, being six years older than Einstein, Besso could not wait too long before embarking on a thesis. Yet it was exactly in 1900, when Einstein was pushing him to become '*Dozen'* – and this required a thesis – that Michele's situation grew more stable. But Michele's father died in October 1901 and Michele moved to Trieste. This would imply that it is likely that Michele, urged by Einstein, did embark on a thesis in autumn 1900, and gave up on it one year later.

### 4.4 Albert Einstein and the *Istituto Lombardo* library.

The letters Einstein sent to Mileva indicate he was in Milan three times a year: in September (before classes resumed at ETH Zürich), at Christmas, then from mid-March until the end of April (semester break). So Einstein would spend approximately three months a year there, pursuing his reflections on science and bibliographic research. On 10th April 1901, he wrote to Mileva from Milan: *"Last week I studied electrochemistry and chemical reactions*

---

[49] This book can only be found in the Milan Observatory in Brera, in a library that Einstein apparently did not visit as he showed no interest for astronomical subjects in his letters to Mileva. Besides, it did not own the physics journals he was interested in.

[50] This might be the reason why the books on wireless telegraphy can be found among the few books in physics in Jung's library: A. Broca (*La télégraphie sans fils*, Paris, Gauthier-Villars, 1899), H. Poincaré (*La théorie de Maxwell et les oscillations hertziennes*, Paris, Carré et Naud, 1899) and R.Colson (*Traité élémentaire d'électricité,* 1900), which mentions this subject at the end of this third edition.

*from Michele's 'Ostwald', and the electron theory of metals in the library"* (*CP* 1 Doc. 97). Therefore, a natural question that arises, and which to our knowledge has not been examined, is to know in which library Einstein was working in Milan.

Some evidence can be found from these letters:

1) On 28th September 1899, concerning the relative motion: *" I too have done much bookworming & puzzling out, which was in part very interesting. I also wrote to Professor Wien in Aachen about the paper on the relative motion of the luminiferous ether against ponderable matter, which the 'Principal' treated in such a stepmotherly fashion. I read a very interesting paper published by this man on the same topic in 1898* " (*CP* 1 Doc. 57). Thus this library must have had the *Annalen der Physik und Chemie* in which was published (Wien 1898).

2) On 4th April 1901, concerning Drude: "*On the other hand, I have in my hands a study by Paul Drude on the electron theory, which is written to my heart's design, even though it contains some very sloppy things. Drude is a man of genius, there is no doubt about that. He also assumes that it is mainly the negative electric nuclei without ponderable mass which determine the thermal and electric phenomena in metals, exactly as it occurred to me shortly before my departure from Zürich*" (*CP* 1 Doc. 96). Thus in Milan Einstein would work on the papers (Drude 1900a; 1900b) (as indicated in the CPAE) which appeared in the *Annalen der Physik*, which replaced the *Annalen der Physik und Chemie* in 1900.

3) On 10th April, after expressing doubts regarding Planck's resonators: *"It is easy to explain what is setting me against Planck's considerations on the nature of radiation, an assumption I cannot really warm up to. Maybe his newest theory is more general. I intend to have a go at it"* (*CP* 1 Doc. 97). Even if the communication delivered by Planck on 15th December 1900 (Planck 1900) before the German Physical Society, when he presented the quantization of radiation energy on Hertzian Resonators, was not available in Milan, it would be reported in the *Beiblätter zu den Annalen der Physik* in March 1901 (Planck 1901a) and published in the March issue of the *Annalen* (Planck 1901b). As it seems Einstein was aware of the existence of this theory, it is likely this library also owned the *Beiblätter*.

4) On 4th April, again: "*Now I need to go to the library or it'll be too late*". This sentence not only confirms the fact that he would visit a certain library in particular, but also that it was close to where he lived (which does not mean it could be a discriminating criterion).

At that time it was only the library of the *Istituto Lombardo, Accademia di Scienze e Lettere* in Milan that owned these journals. [51] It was located in the *Brera* Palace five hundred meters from Einstein's home. It should be noted that the *Istituto Lombardo* held a very rich collection of periodicals, but lacked academic reference works in German. For instance, Ostwald's publication was not available (which might explain why Einstein borrowed it from Besso), nor were Helmholtz's, Boltzmann's or Mach's publications and Einstein intended to have them sent to him from Zürich to Milan in September 1899 (*CP* 1 Doc. 54). There is thus a strong correlation between the *Istituto Lombardo* library's collection - a library Einstein could have access to thanks to Michele's family for instance, and the information found in the letters sent to Mileva.

Let us recall that the *Istituto Lombardo* was founded following the French model of the *Institut de France* by Napoleon Bonaparte in 1797 *"who wanted for the Cisalpine Republic* an '*Istituto Nazionale', entrusted to collect discoveries and improve arts and*

[51] This library is today located at 25 via Borgonuovo, in the Institute's headquarters, in the Palace *Landriani*, opposite the *Brera* Palace entrance.

*sciences"* (Robbiati 2013).[52] The *Brera* Palace houses the *Istituto Lombardo* and the *Braidense* library too (and its one million books which date back to the 15th century), the Observatory of Milan (and its library of 30,000 volumes), the Botanical Garden, the *Accademia di Belle Arti di Brera,* and its famous Picture Gallery, founded by Napoleon too. We appreciate better now a quote from Reiser who summed up the years Einstein spent in Milan as follows: *"Milan was a paradise of freedom and beauty. He read much now, with that complete passion and devotion with which young men read. But he also enjoyed the landscape of Northern Italy, and the sun and warmth, which he loved about all things. For the first time in his life, he studied the plastic and graphic arts: the last supper of Leonardo da Vinci at Santa Maria delle Grazie, the collections at Brera – a world of classical beauty!"* (Reiser 1931, p 42). If Einstein showed an interest in the art collection at *Brera,* however it was not the main link he had established with the *Brera* Palace.

### 4.5 The shift in the orientation of Albert Einstein's thesis in April 1901.

We can therefore approach Einstein's work in Milan by bringing together the scientific information provided by his letters to Mileva with the collection of the *Istituto Lombardo* library. In particular, this allows a better understanding of the reason why the submission of his initial thesis on molecular forces was delayed and reoriented in April 1901. In December 1900, Mileva was indeed expecting the submission of Albert's manuscript for Easter 1901: *"Albert is still here* [in Zürich] *and is going to stay until he finishes his doctoral thesis, which will probably take until Easter"* (CP 1 Doc. 85). But in April 1901, from Milan, Einstein told his friend Marcel Grossmann he was extending his subject to low compression gas. In a letter of 14th April, he wrote: *"As for science, I have a few splendid-ideas, which only now need proper incubation. I am now convinced that my theory of atomic attraction forces can also be extended to gases* […]. *That will also bring the problem of the inner kinship between molecular forces and Newtonian action-at-a-distance forces much nearer to its solution. It is possible that experiments already done by others for other purposes will suffice for the testing of the theory. In that case I shall utilize the already existing results in my doctoral dissertation. It is a glorious feeling to perceive the unity of a complex of phenomena which appear as completely separate entities to direct sensory observation"* (*CP* 1 Doc. 100). On the next day, he wrote to Mileva in similar terms: *"As for science, I've got an extremely lucky idea, which will make it possible to apply our theory of molecular forces to gases as well* […]. *I can hardly await the outcome of this investigation. If it leads to something, we will know almost as much about molecular forces as about gravitational forces, and only the law of radius will still remain unknown*" (*CP* 1 Doc. 101).

Albert Einstein, who was then 22, and doing bibliographic research at the *Istituto Lombardo,* could refer to Maximilian Reinganum's paper entitled *On molecular forces in low compression gas* (Reinganum 1900), which could account for the reorientation of his thesis subject. Indeed Reinganum was considering the dissociation of molecules by taking into account 'planetary' *gravitational forces* in $r^{-4}$. In his opinion, the molecular size played no role in the dissociation processes (contradicting Boltzmann) and he considered them as centers of force. Einstein shared this view of the matter and reminded it to Mileva on 30th April 1901: *"I am very curious whether our conservative molecular forces will hold good for gases as well. If the mathematically so unclear concept of molecular size does not show up as well in the formation of trajectories of molecules coming close to each other as well, but the molecule can be conceived as center of force. We shall get quite a precise test of our view*" (*CP* 1 Doc. 102). Reinganum's paper was published in the *Festschrift* of 11th December 1900, which commemorated the 25th anniversary of Lorentz's doctoral dissertation, and many leading scientists gathered to celebrate the event in Leyden. These communications were published in a volume of the *Archives néerlandaises des sciences exactes et naturelles,* the *Istituto Lombardo* library received on 31st January 1901. [53] Besides, Reinganum was an author Eisntein knew already for he had certainly seen him quoted in Boltzmann's work. He was to read another paper by Reinganum the next month, in May 1901, in Winthertur, about an issue of great importance for him: Drude's classical free electron theory (*CP* 1 p 237; Doc. 111, n 9).

[52] At the origin, the Istituto Lombardo was divided into three classes: Physical Sciences and Mathematics, Moral and Political Sciences, Literature and Plastic Arts. Its first president was Alessandro Volta in 1804.

This reorientation of Einstein's thesis would prove effective, as expressed in a letter by Mileva to her friend Helena Savić, in November or December 1901: *"Albert has written a magnificient study, which he submitted as his dissertation. He will probably get his doctorate in a few months* […]. *It deals with the investigation of the molecular forces in gases using various phenomena"* (*CP* 1 Doc. 125). [54]

## 5. Conclusion

The bond between Albert Einstein's and Michele Besso's families came from the Polytechnic school environment from which Albert and Michele as well as their respective uncles Jakob Einstein and Vittorio Cantoni all graduated, and from the electric power industry in Italy. The meeting between Albert Einstein and Michele Besso has thus been put back in its right context. If they met before the autumn 1896, Michele may have been instrumental in Albert's initial training background.

The Einstein family moved to Milan and Pavia as a result of contacts already established a long time before with Italy. In this fast-developing industrial landscape undergoing fundamental restructuring processes, the Einsteins were in contact with leading figures from the industrial electricity sector and the financial sector (Colombo, Ferraris, Luzzatti, etc.). These leading figures who were also related to Michele Besso's family, certainly played a role in enabling the Einstein family to settle under excellent conditions, despite the difficulties of the family business.

Albert and Michele were reunited in Milan from 1899 to 1901. This period is important to understand Einstein's training background. He carried on a research project that he had undertaken in Zürich and would discuss it regularly with Michele. It foreshadowed the informal collaboration they would engage in as from 1904 in Bern.

The *Istituto Lombardo, Accademia di Scienze e Lettere,* which provided the link between the industrial sector and scientific research, was a major player in the economic development of Lombardy. There were many members of the *Istituto* who taught at the *Politecnico,* for instance Michele's uncle, Giuseppe Jung with whom Einstein seemingly remained in contact after he left for Bern.

In Milan, Einstein was working on papers published in the scientific journals available at the *Istituto Lombardo* library located in the *Brera* Palace near his home. A crossed reading of the scientific observations Einstein sent to Mileva Marić with this library collection, sheds new light on that time of Albert Einstein's life, as for example, the shift in the orientation of his thesis.

As a conclusion, it should be noted that in Milan, Einstein enjoyed an extremely favourable scientific environment and very good working conditions which allowed him to develop his ideas. It is therefore important to deepen our knowledge of this Milanese period to understand better the development of his questioning mind.

---

[53] The *Festchrift* in honor of Lorentz also appeared in the mathematics library of the Milan *Politecnico* which at that time did not own the *Archives Néerlandaises des Sciences Exactes et Naturelles (Deutch Archives of Natural Sciences).* It seems that this copy of the *Festschrift* belonged to Max Abraham, professor at the *Politecnico* from 1909 to 1915.

[54] It was not submitted and Einstein obtained the refund of his registration fees from the University of Zürich on 1st February 1902 (*CP* 1 Doc. 132).

**Acknowledgements:** I would like to thank most sincerely Professor Gianpiero Sironi, President of the *Istituto Lombardo, Accademia di Scienze e Lettere*, for his welcome and for having made himself available for our interviews, as well as the employees of the *Istituto*. I am most grateful to Professor Andrea Silvestri, for proofreading this text and for his most useful comments to clarify several points. I want to thank Valeria Cantoni, Tullo's great-granddaughter, and Vittorio Cantoni, Vittorio's grandson, for providing me with information on their family. I also wish to extend my thanks to Dottoressa Giuseppina Colombo, administrative manager of the *Brioschi* mathematics library at the Milan *Politecnico*, for her cooperation and Mr Andrea Sottile, from the *Ricci* mathematics library at the University of Milan, for his decisive help since the beginning. I also thank Matelda Lo Fiego for her documentary support, my colleague and friend Jean-Pierre Provost for our discussions, as well as my father for assisting me with the Italian translation. Lastly, my sincere gratitude goes to Claudine Dumaine for translating in English the original French text.